# An improvement of the SBU-YLIN microphysics scheme in squall line simulation

Qifeng QIAN*[1], and Yanluan Lin[1]

**ABSTRACT**

The default SBU-YLIN scheme in Weather Research and Forecasting Model (WRF) is proved having a limited capability of producing a reasonable cold pool in squall line simulations. With the help of wealthy observation data of a squall line, we finally improve it by: adding a density factor to the precipitating ice; modify the rain evaporation scheme and correct the saturation adjustment process. The improved SBU-YLIN scheme could produce a reasonable cold pool. Simulation of another two squall line cases in northern and eastern China confirm our modifications are effective. In addition, further analysis reveals that the modification of rain evaporation scheme is the most important among them.

**Key words:** WRF, SBU-YLIN, cold pool, squall line.

---

*Corresponding author : Qifeng Qian
Email: qqf1403321992@gmail.com

# 1. Introduction

The Stony Brook University microphysics scheme (SBU-YLIN) is a single moment bulk microphysics scheme (BMP) which is initially developed in the context of wintertime orographic precipitation (LIN, 2011). It defines a new kind hydrometer category called precipitating ice instead of snow and graupel. The precipitating ice is assumed to be a function of temperature and riming intensity. The latter, which is non-dimensional, is defined as the ratio of the riming growth rate to the riming plus deposition growth rate. Such a diagnostic approach is computationally efficient because it does not require additional prognostic variables. However, the disadvantage is that particle properties are calculated locally and are not tracked in time and space. The mass-size and area-size relationship is a function of temperature when riming does not happen (Heymsfield, 2007), and these are modified through increasing roundness and density of particles during riming. The fall speed of precipitating ice is calculated through Best number method (Mitchell, 1996). The precipitating ice is assumed to follow the inverse-exponential size distribution. The SBU-YLIN scheme also adopts the cloud-rain auto-conversion scheme developed by Liu and Daum (2004).

Although previous studies show that the SBU-YLIN scheme preforms quite well in orographic precipitation simulation (LIN, 2011; Morrison 2015), it shows a bad performance in squall line simulation due to the incapability of producing a reasonable cold pool (Morrsion 2015). One possible explanation for this phenomena is that the fall speed of precipitating ice is too small in squall line simulation. As a result, there is limited precipitating ice falling into the melting layer in the SBU-YLIN scheme, leading to a very weak cold pool that could not be able to initiate secondary convection and maintain the squall line (Morrison, 2015).

In this study, we simulate a squall line on 22 May, 2014 which is well measured by the instruments of the Southern China Monsoon Rainfall Experiment (SCMREX) (Luo, 2016; Qian, 2016). With the wealthy observation, we finally improve the capability of SBU-YLIN scheme in squall line simulation. Simulations of another two documented squall line cases happened in northern and eastern China (Tao, 2014; Zhang, 2015) guarantee the effectiveness of our improvement and provide materials for further analysis.

The paper is organized as follows: section 2 is case description and model setup; the improvement and analysis of our simulation would be presented in section 3; section 4 is the conclusion.

# 2. Case description and model setup

A squall line passed Guangdong Province during SCMREX and caused extreme rainfall event in Qingyuan. The maximum precipitation was as high as 60mm/h. Detailed analysis in previous study shows when the squall line passes, the surface temperature would decrease at most -3K (Qian, 2016). They use the decreasing of surface temperature as the intensity of cold pool, thus, the observed cold pool is -3K. Numerical simulations reveal that SBU-YLIN could not even produce a cold pool in this case (Qian, 2016). In this study, we denote simulation of this squall line as IMPROVE.

Another squall line happened in the northern China caused lightening event and is well analyzed by Zhang et al. (Zhang et al., 2016). It passed Beijing, Tianjin and Hebei and produced strong precipitation. We denote simulation of this squall line as Test01.

The last squall line, happened in Shanghai and northern Zhejiang Province, was as long as 300km and accompanied with hail (Tao, 2014). We denote simulation of this squall line as Test02.

Numerical experiments are performed using WRF v3.6.1 (Skamarock, 2005) with three domains (Fig.1, 27km-9km-3km, 3km is used in all analysis in this study) and two-way nesting for all three cases. 31 sigma levels from the surface to 50hPa are used in the vertical. Model initial and lateral boundary conditions are from the National Center for Environment Prediction Final Operational Global Analysis data (NCEP FNL), available at 1° x1° horizontal resolution and 6-hour interval. Beside SBU-YLIN BMP, other parameterization schemes including: Kain-Fritsch convective scheme (Kain, 1999), RRTMG long and short wave radiation scheme (Iacono, 2006), YSU boundary layer scheme (Hong, 2006) and Noah land surface scheme (Chen, 2001) are used for all three cases. The convective scheme is shut down in 3km domain. A 24-hour simulation of IMPROVE starts from 12:00 UTC, 21May, 2014; Test01 starts from 00:00 UTC, 31 July, 2013; and Test02 starts from 18:00 UTC, 28 March, 2014.

# 3. Improvement and Analysis

IMPROVE shows that the default SBU-YLIN scheme could not even produce a cold pool at all (Fig.2a). The weak cold pool could not trigger secondary convection and thus SBU-YLIN scheme could not maintain the squall line, which leads to a bad performance in squall line simulation and precipitation simulation. Morrison et al. pointed out that fall speed of precipitating ice is too small, which should be blamed for the weak cold pool.

Density factor defined as:

$$density\ \ factor=\sqrt{\frac{1.29}{\rho_a}}$$

could help the precipitating ice fall faster ($\rho_a$ is air density). Fig.2b shows that after adding a density factor to the precipitating ice, the SBU-YLIN scheme could produce a -1K cold pool. However, this weak cold pool forms and dissipates very fast, we could only find it appear at several times (not shown), which indicates that the small fall speed of precipitating ice is not the only problem relating to weak cold pool.

Further analysis reveals that rain droplets stop evaporating when relative humidity is larger than 90% (not shown). The rain evaporation is one of the major sources of latent cooling in low levels, which plays an essential role in the formation and maintenance of cold pool (Qian, 2016). This implies that the rain evaporation scheme could be another possible reason leading to the weak cold pool. In SBU-YLIN scheme, the rain evaporation rate is controlled using following formulas:

$$min\left\{0,(\frac{q_v}{q_{sw}}-0.9)*q_{sw}\right\}$$

where $q_v$ is the water mixing ratio and $q_{sw}$ is the saturation mixing ratio on water surface. Controlling the rain evaporation rate when relative humidity is larger than 90% is a useful technique when the time step is large. This is because calculation of rain evaporation rate is based on the conditions at the beginning of each time step. Applying the initial rain evaporation rate for the whole time step could cause enough vapor to be produced by evaporation that the grid cell would be super saturated when the time step is large. In order to overcome this nonphysical phenomena, SBU-YLIN scheme allow rain evaporation to increase the relative humidity within the grid cell to 90% at the end of the time step based on the initial saturation mixing ratio at each time step. As rain evaporation cools the air and reduces the saturation mixing ratio, 90% threshold of relative humidity provides a marginal safety (Peter Blossey, personal communication). However, this threshold probably causes the weak cold pool. After applying the following formulas (Peter Blossey, personal communication):

$$min\left\{0, 0.9*\frac{q_v+q_l-q_{sw}}{1+\frac{L_v}{C_p}\frac{dq_{sat_liq}}{dT}}\right\}$$

$$\frac{dq_{sat_liq}}{dT}=\frac{q_{sw}L_v}{R_v T^2}$$

$q_l$ is cloud water mixing ratio, $L_v$ is the latent heat of vaporization, $C_p$ is the heat capacity at constant pressure for dry air (J/(kg·K)), $R_v$ is the gas constant for water vapor (J/(kg·K)). These formulas calculate the amount of evaporation that would produce saturation (accounting for the change in saturation mixing ratio due to evaporation cooling) and then make sure that the evaporation would always be smaller than 90% of the amount. After this correction, a very small -2K cold pool is produced (Fig.2c). As the density factor could only produce a -1K cold pool occasionally, this -2K cold pool is mainly produced by the modification of rain evaporation.

As the observed cold pool is -3K, there are still some problems exist in SBU-YLIN scheme so far. We notice that the temperature in the convective cells is 2-3 K higher and relative humidity is 4-5% lower than the observation (not shown) and this problem finally traced to the saturation adjustment process of SBU-YLIN scheme. The saturation adjustment scheme works as following steps in default SBU-YLIN scheme: before the saturation adjustment, the SBU-YLIN scheme first check whether this grid cell is cloudy or not; if the grid cell is cloud free, the temperature at the beginning of the time step is used to calculate the saturation mixing ratio, while if the grid cell is cloudy, it is unsuitable to use this temperature. The true saturation mixing ratio of a cloudy grid cell would differ from the cloud free saturation mixing ratio. The default SBU-YLIN scheme uses the temperature at the beginning of the time step to compute the saturation mixing ratio, this saturation mixing ratio would be a little larger than the true value. As a result, a cloudy grid cell would sometimes become cloud free and then become cloudy at the next time step because the computation of the saturation mixing ratio was correct in cloud free air but not in cloudy air. In fact, the SBU-YLIN scheme should first compute the temperature that would be achieved if the cloud evaporated/sublimated and use that temperature to compute the cloud free saturation mixing ratio. As the temperature would decrease when liquid evaporates or ice sublimates, this cloud free saturation mass mixing ratio will be smaller than that computed from the temperature of the cloudy grid cell. Based on these analysis, we adopt the following formula to compute the temperature when all the liquid and ice become vapor (Peter Blossey, personal communication):

$$T_{modified}=T-\frac{L_v}{C_p}q_l-\left(\frac{L_v}{C_p}+\frac{L_f}{C_p}\right)q_i$$

$L_f$ is the latent heat of melting (J/kg), other variables are same to the rain evaporation scheme. We should emphasize that this formula is only used to determine whether the grid cell is cloudy or not, it would not change the temperature of air during calculations. After fixing this problem, the SBU-YLIN scheme successfully produces a -3K cold pool and well organized convection (Fig.2d). The modification of the saturation adjustment process contributes a -1K cold pool, however, this modification is not always effective (Peter Blossey, personal communication). This means that in other cases, the modification of saturation adjustment probably has trivial improvement just like the density factor.

Although we fix these three major problems and successfully make the SBU-YLIN scheme produce a -3K cold pool, it is still not very convincible that these modifications would also be effective in other cases. So we choose another two cases, Test01 and Test02, which help us to make sure that these modifications are effective. Fig.3a-b show that these modifications could improve the cold pool by -1K in Test01 and Fig3c-d show that these modifications also improve the cold pool by -1K in Test02. Test01, Test02 and IMPROVE together confirm that these modifications are actually effective, which could indeed improve the simulation of cold pool. Further analysis show that the modification of rain evaporation scheme contributes the largest part of -1K temperature decreasing in both Test01 and Test02, which is in consistent with what we mentioned before (not shown).

# 4. Conclusions

The default SBU-YLIN scheme could not produce a reasonable cold pool. Through investigating the model outputs carefully, we finally improve the SBU-YLIN scheme by: applying a density factor to the precipitating ice; modifying the rain evaporation scheme which does not allow rain droplets evaporating when relative humidity is larger than 90% in default SBU-YLIN scheme; and correcting the unsuitable temperature which is used to determine whether a grid is cloudy or not in the saturation adjustment process of the default SBU-YLIN scheme. Three squall line cases spreading in southern (IMPROVE), northern (Test01) and eastern (Test02) China together prove that our modifications are effective, which could enhance the cold pool intensity by -3K in IMPROVE, -1K in both Test01 and Test02. Among the three modifications, adding a density factor could only enhancing the cold pool by -1K occasionally in IMPROVE, which is the most trivial improvement. Rain evaporation scheme modification contributes up to -2K to the -3K temperature decreasing in IMPROVE, which is the major improvement among the three modifications. The last modification (saturation adjustment process) is case sensitive, which could contribute -1K in IMPROVE, but has limited effects in Test01 and Test02.

After our modification, the improved SBU-YLIN scheme has a better performance in cold pool simulation. We would submit our modifications to WRF later.

***Acknowledgments.***

We really appreciate the help from Prof. Peter Blossey, University of Washington, who helps us clarify the problems in the default SBU-YLIN scheme and provides very helpful suggestions in our work.

# Caption:

**Fig. 1.** Simulation domain of IMPROVE (a), Test01 (b) and Test02 (c). The shading represents topography.

**Fig. 2.** The radar reflectivity at 500m AGL (color shaded, unit: dBZ) and the 1-hour 2m temperature decreasing in IMPROVE (model time 06:00 UTC, 22 May, 2014).

**Fig. 3.** Similar to Fig.2 but for Test01 (a-b) and Test02 (c-d) using the default SBU-YLIN scheme and the improved SBU-YLIN.

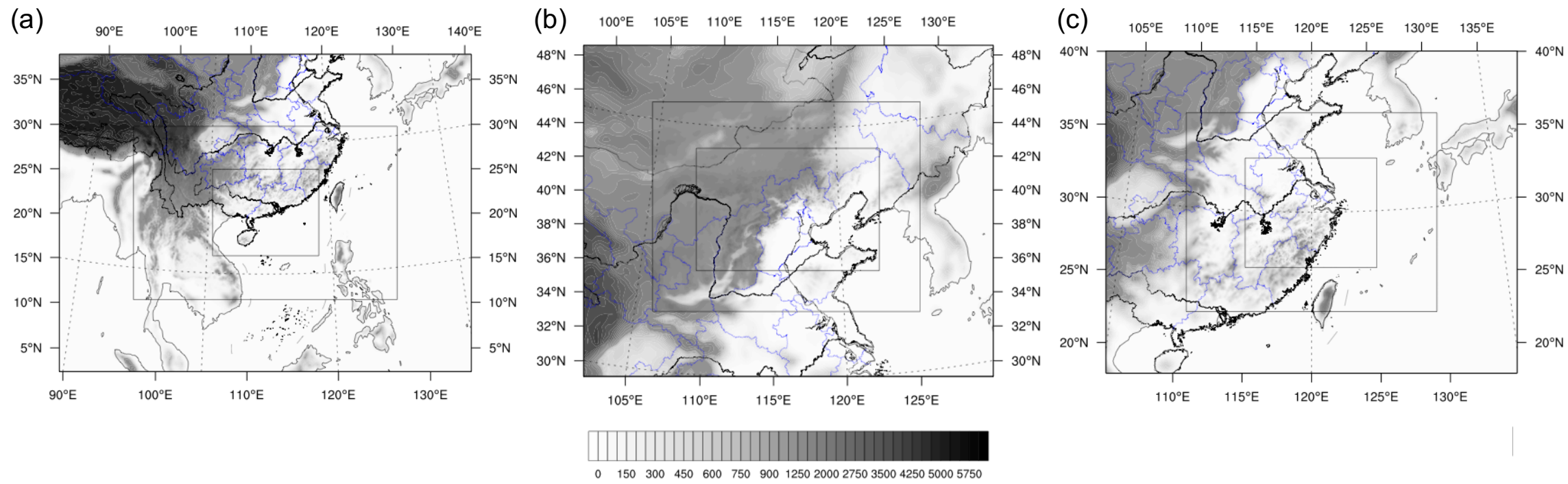

**Fig. 1.** Simulation domain of IMPROVE (a), Test01 (b) and Test02 (c). The shading represents topography.

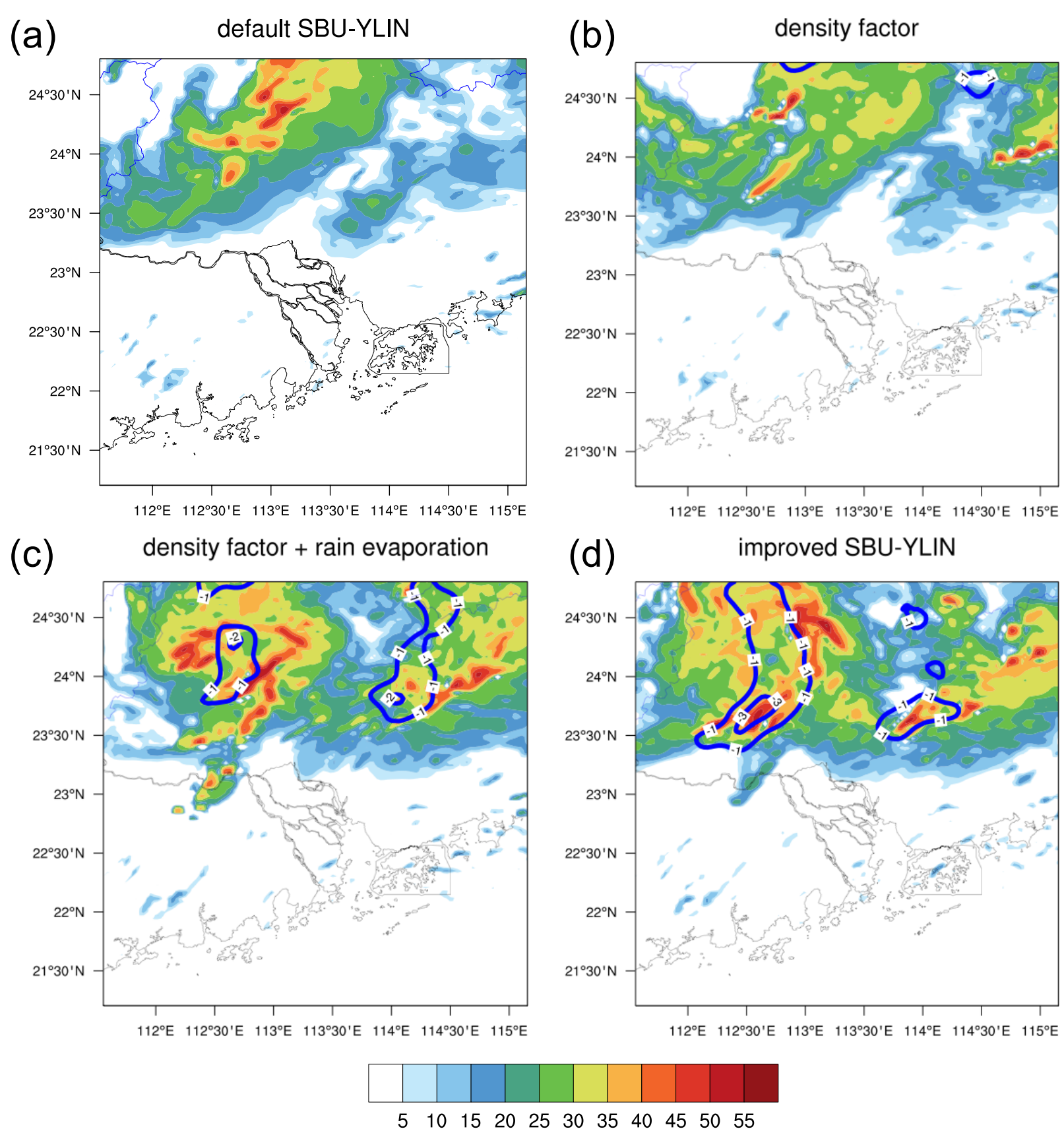

**Fig. 2.** The radar reflectivity at 500m AGL (color shaded, unit: dBZ) and the 1-hour 2m temperature decreasing in IMPROVE (model time 06:00 UTC, 22 May, 2014).

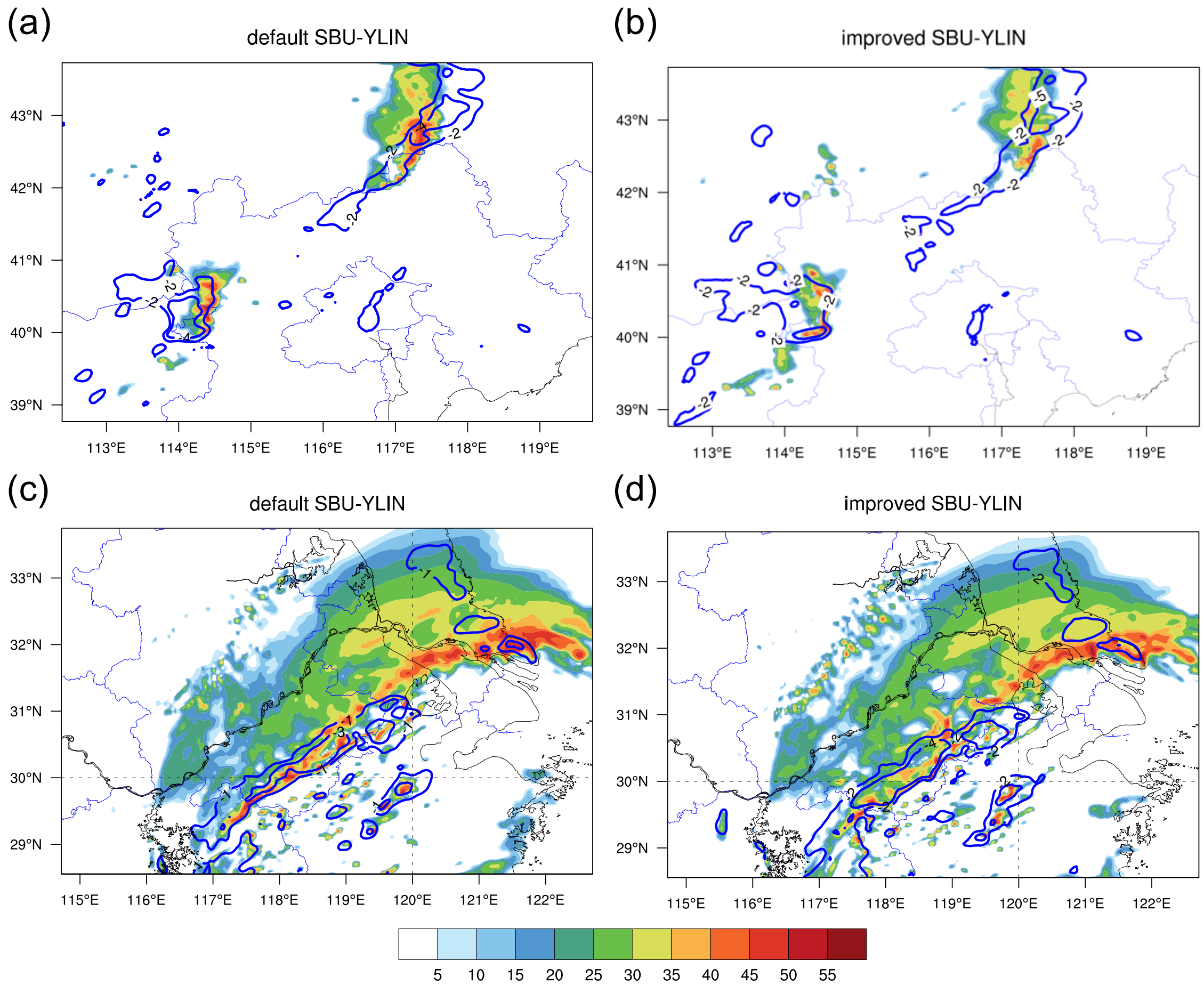

**Fig. 3.** Same as Fig.2. (a-b) is the results of Test01 and (c-d) is the results of Test02 using the default SBU-YLIN scheme and the improved SBU-YLIN.